\documentclass[letterpaper]{article} 
\usepackage{aaai25}  
\usepackage{times}  
\usepackage{helvet}  
\usepackage{courier}  
\usepackage[hyphens]{url}  
\usepackage{graphicx} 
\usepackage{natbib}  
\usepackage{caption} 
\usepackage{algorithm}
\usepackage{soul}
\usepackage[noend]{algorithmic}

\usepackage{newfloat}
\usepackage{listings}
\DeclareCaptionStyle{ruled}{labelfont=normalfont,labelsep=colon,strut=off} 
\floatstyle{ruled}
\newfloat{listing}{tb}{lst}{}
\floatname{listing}{Listing}
\usepackage{enumerate}

\title{Dynamic Symmetry Breaking for Quantified Graph Search: A Comparative Study}
\title{Breaking Symmetries in Quantified Graph Search: A Comparative Study}
\author {
    Mikoláš Janota\textsuperscript{\rm 1},
    Markus Kirchweger\textsuperscript{\rm 2},
    Tomáš Peitl\textsuperscript{\rm 2},
    Stefan Szeider\textsuperscript{\rm 2}
}
\affiliations {
    \textsuperscript{\rm 1} Czech Technical University in Prague \\
    \textsuperscript{\rm 2} Algorithms and Complexity Group, TU Wien, Austria\\
    mikolas.janota@gmail.com, mk@ac.tuwien.ac.at, peitl@ac.tuwein.ac.at, sz@ac.tuwien.ac.at
}

\usepackage{booktabs}
\usepackage{amsthm,amssymb,amsmath}

\usepackage{tikz}
\usetikzlibrary{arrows,matrix,positioning}
\usetikzlibrary{fit}
\usetikzlibrary{arrows.meta,arrows}
\usetikzlibrary{decorations.pathmorphing}
\usetikzlibrary{calc}

\usepackage{cite}

\newcommand{\GGG}{\mathcal{G}}

\def\hy{\hbox{-}\nobreak\hskip0pt} 

\newcommand{\newSolver}{2Qiss}
\newcommand{\qbfstatic}{Q-static}
\newcommand{\qbfsms}{Q-SMS}
\newcommand{\CNF}{\text{CNF}}
\newcommand{\DNF}{\text{DNF}}

\newcommand{\SB}{\{\,}%
\newcommand{\SM}{\mid}
\newcommand{\SE}{\,\}}%

\newcommand{\conn}[1]{\text{connected}^{#1}}
\newcommand{\bipartite}[1]{\text{bipartite}^{#1}}
\newcommand{\cubic}[1]{\text{cubic}^{#1}}
\newcommand{\girthenc}[2]{\text{girth}^{#1}_{#2}}

\newcommand{\problem}{\noindent \textbf{Problem:~}}
\newcommand{\encoding}{\noindent \textbf{Encoding:~}}

\usepackage{boxedminipage}
\newcommand{\pbDef}[4]{%
	\noindent
	\begin{center}
		\begin{boxedminipage}{0.98 \columnwidth}
			
					\textbf{Task:} #2 \\[2pt]
				\textbf{$\exists$\hy Encoding:} #3\\[2pt]
				\textbf{$\forall$\hy Encoding:~} #4
		\end{boxedminipage}
	\end{center}
}

\definecolor{ijcaired}{HTML}{D22817}
\definecolor{azure}{rgb}{0.0, 0.5, 1.0}
\definecolor{applegreen}{rgb}{0.55, 0.71, 0.0}
\definecolor{auburn}{rgb}{0.43, 0.21, 0.1}
\definecolor{bittersweet}{rgb}{1.0, 0.44, 0.37}
\definecolor{byzantine}{rgb}{0.74, 0.2, 0.64}
\definecolor{darkmagenta}{rgb}{0.55, 0.0, 0.55}
\definecolor{darkelectricblue}{rgb}{0.33, 0.41, 0.47}

\definecolor{colorbrewer0}{HTML}{a6cee3}
\definecolor{colorbrewer1}{HTML}{1f78b4}
\definecolor{colorbrewer2}{HTML}{b2df8a}
\definecolor{colorbrewer3}{HTML}{33a02c}
\definecolor{colorbrewer4}{HTML}{fb9a99}
\definecolor{colorbrewer5}{HTML}{e31a1c}
\definecolor{colorbrewer6}{HTML}{fdbf6f}

\newcommand{\lv}[1]{}
\newcommand{\sv}[1]{#1}  

\newtheorem{lemma}{Lemma}
\newtheorem{conjecture}{Conjecture}

\begin{document}

\maketitle
\begin{abstract}
Graph generation and enumeration problems often require handling equivalent graphs---those that differ only in vertex labeling. We study how to extend SAT Modulo Symmetries (SMS), a framework for eliminating such redundant graphs, to handle more complex constraints. While SMS was originally designed for constraints in propositional logic (in NP), we now extend it to handle quantified Boolean formulas (QBF), allowing for more expressive specifications like non-3-colorability (a coNP-complete property). We develop two approaches: a static QBF encoding and a dynamic method integrating SMS into QBF solvers. Our analysis reveals that while specialized approaches can be faster, QBF-based methods offer easier implementation and formal verification capabilities.

\end{abstract}

\section{Introduction}

Generating and enumerating graphs can become inefficient when the same graph structure appears multiple times with different vertex labels. A graph's isomorphism class contains all such equivalent labelings, of which we can designate one as canonical---typically the one with lexicographically smallest adjacency matrix. SAT Modulo Symmetries \cite[SMS;][]{KirchwegerSzeider24} builds on this idea to eliminate redundant graph generation. It extends a conflict-driven (CDCL) SAT solver with a custom propagator that enforces canonicity: when a partial graph assignment cannot be extended to a canonical solution, the solver backtracks. While static symmetry breaking through additional constraints is also possible, no polynomial-size encoding is known for complete symmetry breaking of general graphs.

The original SMS framework handled constraints expressible in propositional logic. However, many interesting graph properties require quantified constraints. For example, proving a graph is not 3-colorable means showing that no assignment of three colors to vertices avoids adjacent vertices sharing a color - a coNP-complete property. To handle such properties, SMS was extended with co-certificate learning \cite[CCL][]{KirchwegerPeitlSzeider23}. CCL uses two solvers: one generates candidate graphs, while another tries to refute them by finding certificates like valid colorings. When a certificate is found, CCL strengthens the constraints for future candidate generation.

While effective, CCL requires custom implementations for each new graph property - both the refutation algorithm and the constraint strengthening mechanism must be built from scratch. This makes CCL tedious to implement and error-prone for complex properties. Quantified Boolean formulas (QBFs) offer a more systematic approach. A QBF solver can automatically handle the interplay between graph generation and certificate checking based on a formal problem specification. This brings three key benefits: simpler implementation through standard encodings, formal verification through proof generation, and automatic incorporation of future advances in QBF solving technology.

In this paper, we develop this QBF-based approach into a complete framework for quantified graph search problems. Our contributions are:
\begin{enumerate}
\item a complete dynamic symmetry-breaking method Q\hy SMS that extends existing QBF solvers with SMS-style pruning;
\item a complete static symmetry-breaking method Q-static using universal variables to encode graph canonicity;
\item implementations of several quantified graph search problems demonstrating the framework's applicability.
\end{enumerate}



\begin{table}[ht]
  \centering
	\begin{tabular}{@{}lccc@{}}
		\toprule
			        & CCL         & \qbfstatic       &
                                                                   \qbfsms
          \\ \cmidrule(l){3-4}
					& (Kirchw.~et al. \citeyear{KirchwegerPeitlSzeider23}) & \multicolumn{2}{c}{this paper}\\
		\midrule
		speed       & \checkmark  &                  & \checkmark         \\
		proofs      &             & \checkmark       & \checkmark         \\
		ease-of-use &             & \checkmark       & \checkmark         \\
		\bottomrule
	\end{tabular}
	\caption{ Pros and cons of the approaches studied in this
          paper.  CCL is typically the fastest but hard to implement
          for complicated problems.  \qbfstatic\ is straightforward
          to implement (one encoding can be reused) and can be used
          with any QBF solver, but turns out to be significantly slow.
          Encodings for \qbfsms\ are just as easy as with \qbfstatic,
          but run much faster; the only price to pay is the need for a
          specialized SMS-endowed solver (such as those presented in
          this paper).  Both \qbfstatic\ and \qbfsms\
can produce independently verifiable proofs, which, combined with
uniform (and hence less error-prone) problem encodings, provide
stronger trust than the slightly faster CCL.
}
	\label{table:overview}
\end{table}

Table~\ref{table:overview} shows the main attributes of the three paradigms for quantified graph search.
In Section~\ref{sec:qbf-sms}, we explain the implementation details for both \qbfstatic, and \qbfsms, describe the QBF solvers we modified and the production of formal proofs, and give an overview of CCL.

For benchmarking, we selected several classes of quantified graph search
problems composed of applications that appeared in earlier SMS papers and other prominent problems in graph theory.
We describe them in Section~\ref{sec:problems}.

In Section~\ref{sec:results}, we discuss the results of our evaluation.
We conclude that for simpler problems, CCL is usually the fastest, but
it cannot provide formal proofs and is generally more error-prone than the other methods.
\qbfstatic\ is the simplest to implement but scales poorly.
\qbfsms\ strikes the best balance overall, providing good performance, formally verifiable proofs, and the expressiveness of general QBF.

Among the QBF solvers themselves, our results show that our newly implemented 2\hy QBF solver, \newSolver, outperforms the other QBF solvers.
This is a surprise, as our primary motivation for putting SMS into QBF solvers was to obtain a performance boost from using established solvers.


\section{Preliminaries}

For a positive integer~$n$, we write $[n] = \{1,2,\dots,n\}$.
We assume familiarity with fundamental notions of propositional logic~\cite{Prestwich21}. 
Below, we review some basic notions from graph theory.

\paragraph{Graphs.}
All considered graphs are undirected and simple (i.e.,
without parallel edges or self-loops). A \emph{graph} $G$ consists of
set $V(G)$ of vertices and a set $E(G)$ of edges; we denote the edge
between vertices $u,v\in V(G)$ by $uv$ or equivalently $vu$. The \emph{order} of a graph $G$ is the number of its vertices,~$|V(G)|$. We write
$\GGG_n$ to denote the class of all graphs with $V(G) = [n]$.  The \emph{adjacency matrix} of a
graph $G \in \GGG_n$, denoted by $A_G$, is the $n\times n$ matrix where the element at row $v$ and column $u$, denoted by $A_G(v,u)$, is $1$ if $vu \in E$ and
$0$ otherwise. 
\lv{We write  $N_G(v)$  for the \emph{neighborhood} of a vertex $v$ in $G$.}

\paragraph{Coloring.}
A \emph{proper $k$-coloring} of a graph $G$ is a mapping
$c : V(G) \to [k]$ such that $uv \in E(G)$ implies $c(u) \neq c(v)$.
The \emph{chromatic number} of a graph $G$ is the smallest integer
$k$, for which a proper $k$-coloring exists.
A \emph{$k$-edge-coloring} of a graph $G$ is a mapping
$c : E(G) \to [k]$. A $k$-edge-coloring of a graph $G$ is \emph{proper} if incident edges have different colors.

\paragraph{Isomorphisms.}
For a permutation $\pi : [n] \rightarrow [n]$, $\pi(G)$ denotes the graph obtained from $G\in \GGG_n$ by the
permutation $\pi$, where $V(\pi(G)) = V(G) = [n]$ and
$E(\pi(G))=\SB \pi(u)\pi(v)\SM uv \in E(G) \SE$.
Two graphs $G_1,G_2\in \GGG_n$ are \emph{isomorphic} if there is a
permutation $\pi : [n] \rightarrow [n]$ such that $\pi(G_1)=G_2$; in this case $G_2$ is an \emph{isomorphic copy} of $G_1$.
%





\paragraph{Partially defined graphs.}
As defined by \citet{KirchwegerSzeider21},
a \emph{partially defined graph} is a graph $G$ where
$E(G)$ is split into two disjoint sets~$D(G)$ and~$U(G)$.  $D(G)$
contains the \emph{defined} edges, $U(G)$ contains the \emph{undefined} edges.  A
(\emph{fully defined}) graph is a partially defined graph $G$ with
$U(G)=\emptyset$.
A partially defined graph $G$ can be \emph{extended} to a
graph $H$ if  $D(G) \subseteq E(H) \subseteq D(G) \cup
U(G)$.

\paragraph{SAT Modulo Symmetries (SMS).}
%
%

SMS is a framework that augments a CDCL  SAT solver~\cite{FichteHLS23,MarquessilvaLynceMalik21} with a custom propagator that can reason about symmetries, allowing to search modulo isomorphisms for graphs in $\GGG_n$ which satisfy constraints described by a propositional formula.
During search the SMS propagator can trigger additional conflicts on
top of ordinary CDCL and consequently learn \emph{symmetry-breaking clauses}, which exclude isomorphic copies of graphs. More precisely,
only those copies are kept which are lexicographically minimal (\emph{canonical}) when
considering the rows of the adjacency matrix concatenated into a single vector. A key
component is a minimality check, which decides whether a partially
defined graph can be extended to a minimal graph; if it cannot, a
corresponding clause is learned.  For a full description of SMS, we
refer to the original work where the framework was introduced~\cite{KirchwegerSzeider21,KirchwegerSzeider24}.
SMS has been successfully applied to a wide range of combinatorial
problems \cite{FazekasNPKSB23,
KirchwegerScheucherSzeider22,
KirchwegerPeitlSzeider23,
KirchwegerPeitlSzeider23b,
KirchwegerScheucherSzeider23,
ZhangPeitlSzeider24,
ZhangSzeider23}.

\subsection{Quantified Boolean Formulas}

\emph{Quantified Boolean formulas (QBF)} generalize propositional logic with quantification. 
We consider \emph{closed} formulas in \emph{prenex} form, i.e., ones where all quantifiers are in the front in the quantifier \emph{prefix}, and the rest---the \emph{matrix}---is a propositional formula. 
An example of a QBF is 
\[ \exists x \exists y \forall z \big( (x \land \lnot y) \lor z \big) .\]
Given a formula $\phi$ and an assignment $\alpha$ of some variables, then $\phi[\alpha]$ is the formula resulting by replacing $x$ with $\top$ if $\alpha(x) = 1$ and replacing it with $\bot$ if  $\alpha(x) = 0$. 
The semantics of a QBF can be defined recursively.
The formula $\exists x \phi$ is true if $\phi[\{x \mapsto 0\}] \lor  \phi[\{x \mapsto 1\}]$ is true.
The formula $\forall x \phi$ is true if $\phi[\{x \mapsto 0\}] \land \phi[\{x \mapsto 1\}]$ is true.

Let $X = \{x_1,\ldots, x_n\}$.
If $Q \in \{\exists, \forall\}$, we write $ Q X \phi$ for $Q x_1 \ldots Q x_n \phi$.
Further, $\alpha|_{X}$ gives the restriction of the function only to elements in $X$.
A simple CEGAR (counterexample-guided abstraction refinement) algorithm for 2-QBF of the form $\exists X \forall Y \phi$ is given in Algorithm~\ref{alg:CEGARsimple}.
The SAT solver used as a subroutine returns a model of the propositional formula if satisfiable, otherwise NULL.
The idea behind Algorithm~\ref{alg:CEGARsimple} is the basis of many QBF solvers~\cite{JanotaKlieberMarquessilvaClarke12,RabeTentrup15,Hecking2018,cqesto,qfun}.

%
%

\renewcommand{\algorithmicrequire}{\textbf{Input:}}
\renewcommand{\algorithmicensure}{\textbf{Output:}}
\renewcommand{\algorithmicloop}{\textbf{loop forever}}

\begin{algorithm}
	\caption{A CEGAR-based 2-QBF solver}
	\label{alg:CEGARsimple}
	\begin{algorithmic}[1]
		\REQUIRE{A closed 2-QBF $\Phi = \exists X \forall Y \phi$.}
		\ENSURE{True if $\Phi$ is true, false otherwise.} 
		\STATE $\phi' \gets \top$
		\LOOP
			\STATE $\alpha \gets \text{SAT}(\phi')$ \hfill \COMMENT{Get a model of $\phi'$} \label{line:call1}
			\IF{$\alpha = \text{NULL}$}
				\RETURN False \hfill \COMMENT{No model}
			\ENDIF
			\STATE $\beta \gets \text{SAT}(\lnot \phi[\alpha|_{X}])$ \hfill \COMMENT{Compute a counterexample} \label{line:call2}
			\IF{$\beta = \text{NULL}$}
				\RETURN True
			\ENDIF
			\STATE $\phi' \gets \phi' \land \phi[\beta|_Y]$ \hfill \COMMENT{Refine} \label{line:refine}
		\ENDLOOP
	\end{algorithmic}
\end{algorithm}

Algorithm~\ref{alg:CEGARsimple} can be seen as a game. One player tries to assign existential variables to satisfy the matrix~$\phi$; the other tries to find universal counter-moves to falsify~$\phi$.
In response to a successful counter-move~$\beta$, the first player must find a move~$\alpha$ that does not lose to $\beta$ (or any other previous move).

\section{Combining SMS with QBF}
\label{sec:qbf-sms}

In this section we expound the approaches listed in Table~\ref{table:overview}.
We start with the idea of checking the canonical form with universal variables (\qbfstatic), and move on to the integration of SMS in existing QBF solvers.

\subsection{Quantified Static Symmetry Breaking (\qbfstatic)}

While it is unknown whether a polynomially sized complete symmetry break is expressible in propositional logic, it is easy to achieve in QBF.
The idea is to encode that no permutation of the vertices results in a lexicographically smaller adjacency matrix for the graph, using universal variables to represent the permutations.

We first construct a formula expressing that a permutation leads to a lexicographically smaller graph for a fixed number of vertices $n$.
Negating the formula leads to the encoding of minimality.
Since the adjacency matrix is symmetric, it is sufficient to consider only the upper triangle.
We write $P_n:= \SB (i,j) \SM i,j \in [n], i < j \SE$, and use the variables $e_{i,j}$ for $(i,j) \in P_n$ to encode the adjacency matrix and 
$p_{i,j}$ for $i,j \in [n]$ to encode the permutation $\pi: [n] \to [n]$, i.e., $p_{i,j}$ is true if and only if $\pi(i) = j$. 
We use the following formula to ensure that the variables $p_{i,j}$ indeed represent a permutation, i.e., that $\pi$ is total and injective:
\[ \mathit{isPerm} = \bigwedge_{i \in [n]} \bigvee_{j \in [n]} p_{i,j} \land \bigwedge_{i<j} \bigwedge_{k \in [n]} (\lnot p_{i,k} \lor \lnot p_{j,k}). \]

\noindent
Finally, we define variables $\mathit{pe}_{i,j}$ to hold the permuted adjacency matrix\lv{ based on the permutation represented by $p_{i,j}$}:
$\mathit{pe}_{i,j}$ is true if, and only if $A_{\pi(G)}(i,j) = 1$.
\[ \mathit{pe}_{i,j} = \bigvee_{(i',j') \in P_n} (e_{i',j'} \land p_{i',i} \land p_{j',j}) . \]
 
\noindent This allows us to construct a formula that is true if the permutation leads to a lexicographically smaller graph:
\[\mathit{nonMin} = \!\!\! \bigvee_{(i,j) \in P_n} \!\!\! \bigwedge_{\substack{(i',j') \in P_n, \\ (i',j') \prec_{\mathit{lex}} (i,j)}} \!\!\!\!\! (e_{i',j'} \lor \lnot \mathit{pe}_{i',j'}) \land e_{i,j} \land \lnot \mathit{pe}_{i,j} , \]
where $\prec_{\mathit{lex}}$ is the lexicographic order on vertex pairs, i.e., $(i,j) \prec_{\mathit{lex}} (i',j')$ if (i) $i < i'$ or (ii) $i = i'$ and $j < j'$.

\lv{We extract common subexpressions to make the formula more compact.}

If the formula $\mathit{nonMin} \land \mathit{isPerm}$ is satisfied, then there is a permutation $\pi$ described by the assignment to $p_{i,j}$ and $(i^*,j^*) \in P_n$ such that $A_{G}(i',j') \geq A_{\pi(G)}(i',j')$ for all $(i',j') \prec_{\mathit{lex}} (i^*,j^*) $ and $A_{G}(i^*,j^*) > A_{\pi(G)}(i^*,j^*)$. This implies that $G$ is not minimal, and consequently the formula
\[ \forall \SB p_{i,j} \SM  i,j \in [n] \SE \quad  \lnot (\mathit{nonMin} \land \mathit{isPerm}) \]
encodes lexicographic minimality.

As we will see in the experiments, trying to enforce
lexicographic minimality using a QBF encoding scales poorly in terms of performance.
This does not come as a big surprise for CEGAR-based approaches using
a second SAT solver to check whether the universal property is satisfied.
The reason is that there is a ``hidden'' pigeon-hole principle. This is easiest explained in an example. Let $G \in \GGG_{n}$ and let the first vertex have degree $\delta$ with $(1,i) \in E(G)$ for $i \in [n] \setminus [n - \delta]$. Let another vertex $u$ have degree $\delta + 1$. When mapping the vertex $u$ to the first vertex, the solver tries to assign the $\delta + 1$ neighbors of $u$ to vertices in $[n] \setminus [n - \delta]$. In other words, one tries to injectively map $\delta + 1$ elements to $\delta$ elements. SAT solvers are known not to perform well on these types of instances.


Independently of the chosen ordering, a further disadvantage of the
minimality encoding is that, for CEGAR-based approaches, the
minimality is only checked when all existential variables are
assigned, i.e., the graph is fully defined.  One can often detect that
a partially defined graph with only a few edge variables assigned
already cannot be extended to a canonical one, and thus branches of
the search tree can be pruned much earlier.  An extreme case is
provided by instances where the existential part is unsatisfiable: the
static symmetry-breaking constraint is not evaluated at all.

\lv{
Another difference worth noting is the type of constraints which are learned.
In SMS only a single clause is learned to discard a partially defined graph.
In the QBF encoding a permutation is learned, i.e., the formula is strengthened to exclude graphs which are made lexicographically smaller by the learned permutation.
This might be a potential advantage but the size of the constraint is much larger. }

\lv{
An interesting research question could be to determine which QBFs might be well suited for early construction of counterexamples, i.e., before all existential variables have been assigned.
This could for example speed up the solution of instances where the majority of the time is spent for trying to solve the existential part. }

\subsection{QBF Modulo Symmetries (\qbfsms)}
\label{sec:integration}

In contrast to encoding the SMS minimality check into QBF, as
described above, we describe here how to integrate SMS with an external
minimality check (similarly to the SAT-based SMS) into three
different circuit-based QBF solvers, Qfun~\cite{qfun},
CQesto~\cite{cqesto} and Qute~\cite{PeitlSlivovskySzeider19b}.
The main reason for only considering circuit-based solvers is to avoid an additional quantifier alternation resulting from transforming quantified circuits into a QCNF\@.


Qfun and CQesto are both based on CEGAR\@.
Unlike its precursor, the RAReQS algorithm~\cite{JanotaMarquessilva11,JanotaKlieberMarquessilvaClarke12,janota-ai16},
\textbf{Qfun} learns functions using decision trees. Instead of adding
$\phi[\beta|_Y]$ to the formula in Algorithm~\ref{alg:CEGARsimple}
line~\ref{line:refine},  the universal variables are occasionally replaced by formulas/functions only depending on existential variables. For example, a universal variable could be substituted by the negation of an existential variable but also with more complex functions. The functions are constructed based on previous models and counter models.

The \textbf{CQesto} algorithm has been designed to be more lightweight than RAReQS,
which scales poorly beyond 3 levels. Consider
the following example. Let $g\equiv(x\lor y)$ in $\exists x \exists y \forall u\,(g\lor
u)\land(\lnot g\lor\lnot u)$ and the assignment $x=y=1$ to which
the universal player responds with $u=1$, falsifying the second conjunct.
Now, the solver could add the constraint $\lnot x\lor\lnot y$ because
setting both $x$ and $y$ to true is losing for the existential
player.
However, this is a weak constraint because the same move will beat any assignment that satisfies gate~$g$.
 Instead, CQesto \emph{propagates} the
assignment and learns the constraint $\lnot g$.
%
For the purpose of this paper, the solver is run repeatedly to obtain
many different solutions, and we observed that value propagation
incurs a non-negligible time overhead. Therefore, we have changed its
implementation to be bottom-up, instead of top-down, and to evaluate gates
lazily. This has improved the performance slightly, but propagation still takes
considerable time.

Note that the solvers are not restricted to 2-QBF, but for the sake of simplicity, we restricted the presentation to this simpler case. For the generalization to an arbitrary number of quantifier alternations, we refer to the literature.  

From a conceptual standpoint, it is straightforward to incorporate SMS into the two CEGAR-based solvers Qfun and CQesto.
Using an SMS solver for calls in Line~\ref{line:call1} ensures that non-canonical graphs are excluded. For all other SAT calls, a standard SAT solver is used.
%

For graph problems, often not only the existence of a graph with a certain property is of interest but also to enumerate all graphs up to isomorphism with the wished-for property. We also extend the solvers to allow enumerating solutions. We propose to use free variables to indicate which variables are relevant for enumeration, i.e., only the assignments of the free variables are part of the output.
Free variables are already part of the QCIR input format~\cite{JordanKlieberSeidl16}.


\textbf{Qute} is a solver based on Quantified Conflict-Driven Constraint Learning~\cite[QCDCL;][]{ZhangMalik02}.
As such, Qute does not use SAT solvers, so we cannot simply replace one component.
It is reasonably straightforward to call the minimality check at the
appropriate place in the QCDCL loop (when unit propagation reaches a
fixed point). However, the learning of symmetry-breaking clauses poses
some challenges that we try to explain briefly.

QCDCL, as implemented in Qute, maintains, in addition to the usual set of input and learned clauses like in CDCL, another set of \emph{cubes}.
\lv{A cube is just a conjunction of literals, and it is often convenient to think of it as a negated clause.}
\sv{A cube is a conjunction of literals.}
The cube set maintained by Qute is a DNF (disjunctive normal form, i.e., a disjunction of cubes) representation of the input formula, together with further learned cubes (just like the clause set is the original formula plus learned clauses).
\lv{Or equivalently: Qute is solving a formula and its negation at the same time (and the cubes are negated clauses derived from the negated input formula).}

Conjoining a (symmetry-breaking or solution-blocking) clause to the clause set is trivial and fast; one just appends it.
However, conjoining a clause to the cube set, by De Morgan's laws, potentially changes every cube, including learned ones, and that is expensive when done repeatedly.
We can do better if we understand the structure of the cubes.

Suppose we are solving a QBF whose matrix is the circuit $F$.
Qute initializes its cube set with $\DNF(F)$, which is obtained with
Tseitin's well-known translation
procedure~\cite{Tseitin68transl}.\footnote{The translation into DNF
  (Disjunctive Normal Form) is the negation of the better known
  translation into CNF (Conjunctive Normal Form), and uses universal auxiliary variables.}
$\DNF(F)$ consists of two components, $\DNF^*(F)$, which encodes the structure of the circuit as a DNF, and the unit cube $(F_{out})$, which says that the circuit should evaluate to true.
Now, suppose we already have $\DNF(F) = \DNF^*(F) \lor (F_{out})$, and we want to obtain $\DNF(F \land C)$ for a newly added clause $C$.
Observe that much of the structure of the circuits $F$ and $F \land C$ is identical.
It is not too hard to see that in fact $\DNF(F \land C) \equiv \DNF^*(F) \lor \DNF^*(C) \lor (F_{out} \land C_{out})$.
Thus, we can conjoin the clause $C$ by encoding it into DNF, appending to the cube set, and replacing the unit cube $(F_{out})$ with the new output cube $(F_{out} \land C_{out})$.

After this replacement, every cube derived using $(F_{out})$ might be invalid.
Such cubes we call \emph{tainted}, and we extended Qute to keep track of whether a cube is tainted\lv{ with one bit in the constraint data structure}.
After adding a clause, all tainted cubes must be removed.

Three technical aspects of clause addition appear to be important.
First, delete tainted cubes lazily: mark them for deletion, but do not actually clean them from memory (a periodical cleanup is performed by the solver).
Second, when a cube shares a literal with the newly added clause, the cube can be kept even if it is tainted\sv{ (a formal proof of this is an easy exercise; we leave it out to save space)}.
\lv{\begin{lemma}
	\label{lemma:cube-preservation}
	Let $\Phi$ be a QBF, $C$ a clause, $T$ a cube.
	If $\neg \Phi \models \neg T$ and $T \cap C \neq \emptyset$, then $\neg (\Phi \land C) \models \neg T$.
\end{lemma}
\begin{proof}
	Follows since $\left ( \Phi \land C \right ) |_T = \Phi |_T \land C |_T \equiv \Phi |_T $.
\end{proof}}
Third, to add a clause, the solver must backtrack to a consistent state.
\sv{We backtrack all the way; a finer yet complicated analysis might allow keeping some of the search state.}
\lv{
Considering just the added clause, it would be sufficient to backtrack enough to ensure it is not falsified by the assignment after backtrack (and to propagate it if unit).
However, since we invalidate existing cubes, and these could themselves have been responsible for propagation at the time of invalidation, we potentially need to backtrack to undo these propagations as well.
For the sake of simplicity, what we do instead is backtrack all the way whenever a new clause is added.
We leave the possibility of improving the implementation to future work.
}

\subsection{Proofs}

We have implemented basic proof logging with SMS in Qute.
We picked Qute for this as it is the only solver that supports proof logging.
Our proof framework certifies correctness of the obtained solutions (that the graphs have the right properties), and the fact that all solutions have been found (unsatisfiability at the end).
To certify correctness of the symmetry-breaking clauses generated by SMS, we can use the same approach as for propositional SMS~\cite{KirchwegerSzeider24}, as the validity of the symmetry clauses does not depend on the encoding.

Qute outputs proofs in \emph{long-distance Q-resolution}~\cite[LDQ-resolution, for short;][]{ZhangMalik02,BalabanovJiang12}, a clausal proof system (and its counterpart LDQ-consensus which operates dually on cubes).
LDQ-resolution and consensus can prove a QBF in CNF/DNF false and true, respectively.
When solving a QBF with the circuit matrix $F$,
the proof starts from $\CNF(F)$/$\DNF(F)$.
An extra step outside of the proof system, which we have not
implemented, would be necessary to certify the correctness of the circuit-to-CNF/DNF translation.

A single LDQ-resolution/consensus proof can certify one solution or prove that there are no solutions.
In order to certify an enumeration problem, several proofs are needed.
All the necessary proofs can be extracted from the solver trace (the chronological log of all learned clauses and cubes).
We have adapted the extractor \texttt{qrp2rup}~\cite{PeitlSlivovskySzeider18} to handle traces containing multiple proofs.

In a proof trace, all variables and axioms are usually introduced right at the beginning.
In SMS, axioms and auxiliary variables can be introduced on the fly as well.
We adapted \texttt{qrp2rup} to handle such cases.
When new axioms are introduced in the middle of the proof, any tainted cubes must be forgotten (see above).
We do not see how the proof checker could check this without understanding the entire setup, including the circuit structure outside of the clausal proofs.
It appears to us that a dedicated proof system for enumeration would be necessary to correctly capture what happens when a solution is found and blocked for subsequent search.
We leave the design of such a proof system to future work.

On the other hand, \texttt{qrp2rup} can, on top of verifying a proof, extract strategies for the winning player and DRAT proofs for them, and verify them with DRAT-trim~\cite{WetzlerHeuleHunt14}, and can even extract GRAT annotations directly and verify the strategies with the formally verified checker \texttt{gratchk}~\cite{Lammich17}.
We made sure this feature continues to work even with multiple proofs.

\subsection{A New 2-QBF Solver}

Since in preliminary experiments, existing QBF solvers (extended with SMS) performed worse than CCL, we decided to write our own 2-QBF solver \newSolver,
%
%
which  implements Algorithm~\ref{alg:CEGARsimple}. We use
CaDiCaL~\cite{cadical2020} as the underlying SAT solver. We
create two instances of the solver, one for calls at
Line~\ref{line:call1}, the other for calls at
Line~\ref{line:call2}. The first solver is responsible for
the existential part, the second for the universal.  The first
solver is initialized with the formula $\phi \left[ \SB y \mapsto 0 \SM y \in Y\SE \right] $ and the second with  $\lnot \phi$. For the first solver, we use the incremental interface to add additional constraints at Line~\ref{line:refine} (incremental calls preserve learned clauses).
For the second, we use assumptions to fix the assignment to the existential variables in each call.

Like the other QBF solvers mentioned above, we use QCIR as input format, i.e., the matrix of the formula is not restricted to CNF.
Nevertheless, constraints must be transformed into CNF before they are added to the underlying SAT solvers.
This is done using the Tseitin transformation~\cite{Tseitin68transl}.
New variables are introduced for subformulas and constraints added to ensure that the variables are equivalent to the subformulas.
To avoid encoding identical subformulas several times, we use a technique called \emph{gate/structural hashing}~\cite{BalabanovJMS16}. For each solver, a hash map is created which maps each subformula to its corresponding variable.

Before being passed to a solver, each formula is simplified by removing true (false) inputs from conjunctions (disjunctions), replacing empty conjunctions (disjunctions) with true (false), and replacing unary gates by their only input.
\lv{In the future, we plan to exploit additional information from modern SAT solvers to further simplify the formula. For example, one can simplify root level fixed variables. }

Many natural 2-QBF problems can be encoded in the form $\phi = \exists X \forall Y \left( F(X) \land G(X,Y) \right) $, i.e., with a top-level conjunct that uses only existential variables.
After adding $\phi \left[ \SB y \mapsto 0 \SM y \in Y\SE \right] $ to the first SAT solver, we delete the existential part $F(X)$ from the formula (if there are more such existential conjuncts, we remove all of them).

\lv{
A more compact transformation of the matrix into CNF is due to \citet{PlaistedG86}.
It analyzes the polarities of subformulas and reduces the number of clauses when a subformula appears in only one polarity.
We provide an option to use this more compact version.
If selected, gate hashing must also be adapted.
Instead of mapping subformulas to variables, one maps pairs of [subformula, polarity] to variables.
If the variable for the other polarity is already present, it is strengthened instead of introducing a new variable.
A potential drawback of the Plaisted-Greenbaum encoding is that propagation is weaker in the resulting formulas.
}

\lv{
For some applications using a circuit format can be more cumbersome to encode. We still provide the option to encode the problem into two CNFs $F(X)$ and $G(X,Y)$ such that the formula to solve is $\exists X \forall Y (F(X) \land \lnot G(X,Y))$. }

\subsection{Co-certificate learning (CCL)}

\colorlet{gencol}{azure}
\colorlet{colcol}{ijcaired}
\begin{figure}
	\centering
	\begin{tikzpicture}[scale=0.77]
		\node[draw=gencol,thick,rounded corners=2] (CDCL) at (0, 0) {1st block solver};
		\node[draw,rounded corners=2] (fullobj) at (3, 1) { Solution Candidate};
		\node[draw,rounded corners=2] (mincheck) at (0, 2) {Symmetry Check};
		
		\node[draw=colcol,thick,rounded corners=2] (conp) at (7, 0) {2nd block solver};
		
		\node[draw,rounded corners=2] (sol) at (7, 2) {Solution};
		
		\node[draw,dashed,inner sep=2mm,label=below:SMS,fit=(mincheck) (CDCL) (fullobj) (CDCL)] (graphgen) {};
		
		\draw (CDCL)     edge[-latex,bend right=20] (mincheck);
		\draw (mincheck) edge[-latex,bend right=20] (CDCL);
		
		\draw (CDCL) edge[-latex,bend left=10] (fullobj.west);
		\draw (fullobj.east) edge[-latex,bend left=20] (conp);
		\draw (conp) edge[-latex,decoration={zigzag,amplitude=pi/4,segment length=1.5*pi},decorate] node[above] {\small counterexample} node[below] {\small blocking circuit} (CDCL);
		
		\draw (conp) edge[-latex] node[right] {\small UNSAT} (sol);
	\end{tikzpicture}
	\caption{
		CCL.
		CEGAR-based QBF solving with an SMS blackbox follows a similar pattern.
	}
	\label{fig:CCL}
\end{figure}
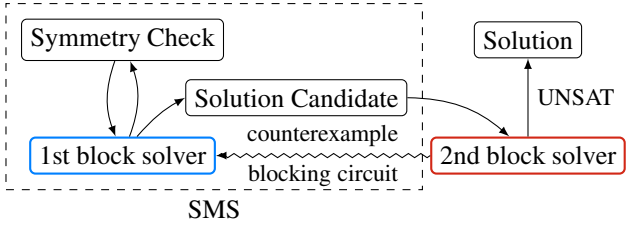

CCL follows the pattern of Algorithm~\ref{alg:CEGARsimple}, depicted graphically in Figure~\ref{fig:CCL}.
The 2nd block solver can be any algorithm: a SAT solver (with a suitable formula), or even a custom domain-specific solver.
Whereas a CEGAR QBF solver computes a strengthening on Line~\ref{line:refine} from the input QBF and a counter-move $\beta$, in CCL, the 2nd-block solver is responsible for returning an appropriate strengthening.
In some cases, like graph coloring, this is easy: when the 2nd solver finds a coloring of the candidate graph proposed by the 1st solver, it can return a clause that says at least one edge with endpoints of the same color should be present in future candidate graphs.
For more involved problems, though, CCL can get complex and error-prone.
Since the 2nd-block solver can be an arbitrary algorithm, it is also hard to provide independently verifiable proofs for CCL.

\section{Benchmark Problems and Encoding}
\label{sec:problems}

In this section we introduce the graph search problems and present QBF encodings on which we evaluated our solvers.
For most problems we only sketch the encoding, as the main focus is on comparing the solving approaches.
See Section~\ref{sec:results} for a link to generator scripts and details of the formulas.
All encodings are of the form $\exists X  \forall Y  ( F(X) \land \lnot H(X,Y))$.
We call $F$ the $\exists$-encoding and  $H$ the $\forall$-encoding.

\subsection{Coloring Triangle-Free Graphs}
\label{sec:triangle-free-def}
If a graph contains the $k$-clique as a
subgraph, then its chromatic number must be at least $k$.  The
opposite is not true: \citet{Mycielski55} explicitly constructed triangle-free graphs (without $K_3$ as a subgraph) with unbounded chromatic number.
Erd\H{o}s~\shortcite{Erdos67} asked about the values $f(k)$, which denote the smallest number of vertices in a triangle-free non-$(k-1)$\hy colorable graph.
Mycielski's construction provides 
upper bounds on $f(k)$, and these are tight up to $k=4$; for $k=5$, minimal graphs are also known~\cite{Goedgebeur20}, but none of them is a \emph{Mycielskian}.
The cases $k \geq 6$ are open.

\newcommand{\col}{c}

\pbDef{(Max\hy)$\triangle$\hy free non-$k$-colorable}
{Compute a triangle-free graph with $n$ vertices and chromatic number at least $k$. }
{The existential part  ensures that the graph is
	triangle-free. 
	Without loss of generality, we further restrict the search to  \emph{maximal triangle-free} graphs (triangle-free and adding any edge creates a triangle). 
}
{Universal variables $\col_{v,i}, v \in [n], i \in [k-1]$ and straightforward constraints enforce non\hy $(k-1)$\hy colorability.}
%

\lv{ We use $c_{i,l}$ for $i \in [n], l \in [k - 1]$ to
indicate the coloring. Let
$X = \SB e_{i,j} \SM i,j \in [n], i < j\SE $ and
$Y = \SB c_{i,l} \SM i \in [n], l \in [k - 1]\SE $.  The existential
encoding is
\[ F(X) =  \bigwedge_{1\leq u < v < w \leq n} (\lnot e_{u,v} \lor \lnot e_{u,w} \lor \lnot e_{v,w})\]
and the universal 
\[G(X,Y) =  \bigwedge_{v \in [n]} \bigvee_{l \in [k - 1]} c_{v,l}     \land \bigwedge_{u<v}\bigwedge_{l \in [k - 1]} (\lnot e_{u,v} \lor \lnot c_{u,l}\lor \lnot c_{v,l}).\] }


\subsection{Folkman Graphs}
\label{sec:folkman-def}

\emph{Folkman graphs}, named after the mathematician Folkman (and
not to be confused with the specific graph also named after
him), play an important role in generalized Ramsey
theory~\cite{Folkman70}.  The \emph{Ramsey number} $R(x,y)$ is the least integer such
that for every $2$-edge-coloring (red-blue) of
the complete graph $K_{R(x,y)}$, one can find a red $K_x$ or a blue $K_y$ subgraph~\cite{Ramsey30,Radziszowski2021}.
Folkman numbers generalize this idea, by looking for the existence of
monochromatic complete subgraphs in edge colorings of arbitrary graphs. 
The \emph{Folkman number}
$F(x,y;k)$ is the least integer such that there exists an $(x,y;k)$\hy \emph{Folkman
  graph}: a $K_k$-free graph such that any
$2$-edge-coloring contains either a red $K_x$ or a blue $K_y$.

\pbDef{Folkman $(3,3;k)$}
{Given two integers $k$ and $n$, output a $(3,3;k)$\hy Folkman graph with $n$ vertices.}
{$K_k$-free: enumerate all $k$-tuples of vertices, and in each require at least one edge to be missing.}
{
	We use universal variables for encoding edge\hy colorings, and require that for each edge coloring there be some monochromatic triangle.}
We focus on the Folkman number $F(3,3;4)$, for which the best bounds are $21 \leq F(3,3;4) \leq 786$~\cite{Bikov20,Lange2012}.
\lv{Note that there is a potential candidate graph with 127 vertices which doesn't contain a 4-clique, but it is not known whether it is possible to find a 2-edge-coloring without a monochromatic triangle.}
%
\lv{ 
It is known that a smallest Folkman graph is not a Sperner graph, i.e.,  $ N_G(i) \not
\subseteq N_G(j)$ for $i,j \in [n], i \not = j$~\cite{Bikov20}. This allows to further restrict the search space.
In addition, we only consider graphs where inserting any additional edge results in a 5-clique. }
%


\subsection{Domination Number of Cubic Graphs}

\newcommand{\dombnd}{\left \lceil |V(G)| /3 \right \rceil}

A \emph{dominating set} of a graph $G$ is a subset $S \subseteq V(G)$
such that each vertex is in $S$ or has a neighbor in~$S$.  The
\emph{domination number} $\gamma(G)$ is the size of a smallest
dominating set of~$G$.
Reed~\shortcite{Reed96} conjectured that the domination number of
every \emph{cubic} (each vertex has degree $3$) connected graph~$G$ is $\leq
\dombnd$. 
This conjecture turned out to be false~\cite{KostochkaStodolsky05,Kelmans2006}, but restricted variants remain open.
A graph is \emph{$k$-connected} if it cannot
be made disconnected by removing fewer than~$k$ vertices.
The \emph{girth} of a graph is the length of its shortest cycle.
\begin{conjecture}
	\label{conj:domination}
	Let $G$ be a cubic graph. If
	\begin{enumerate}[i)]
		\item $G$ is 3-connected, or
		\item $G$ is  bipartite, or
		\item $G$ has girth $g \geq 6$,
	\end{enumerate}
	then $ \gamma(G) \le \dombnd $.
\end{conjecture}
\lv{
Condition (ii) restricts the scope to bipartite (cubic) graphs, but drops the connectivity requirement from condition (i).
It is known that if the girth is at least $83$, then Reed's original
conjecture holds~\cite{LowensteinR08}, and Verstra{\"{e}}te~\cite{Dorbec2024} conjectured that the same holds even under girth $\geq 6$ (condition iii). }

\pbDef{Domination Conjecture}
{Find a counterexample to Conjecture~\ref{conj:domination} with $n$ vertices.}
{We use sequential counters~\cite{Sinz05} to enforce cubicity.
Instead of a full encoding of 3\hy connectedness, we only require
connectedness, and postprocess the solutions.
For girth $g \geq 6$ we use a compact encoding $\girthenc{n}{6}$ due to \citet{KirchwegerSzeider24}.
To encode bipartiteness, we introduce variables $b_i, i \in [n]$ to guess the bipartition.}
{Checking whether a graph does not have a dominating set of size $\leq k$ is coNP-complete.
Universal variables $d_i$, $i \in [n]$ and straightforward constraints describe a dominating set of size $\leq k$.}
\citet{LowensteinR08} proved Reed's conjecture for graphs of girth $\geq 83$;
Verstra{\"{e}}te~\cite{Dorbec2024} conjectured it holds for girth $\geq 6$ (condition \emph{iii}). 

\subsection{Treewidth}
\label{sec:tw-def}
\newcommand{\velim}[2]{#1^*_{#2}}
\emph{Treewidth} is a well-studied graph invariant that measures
how much a graph resembles a tree.  The standard definition of
treewidth uses a \emph{tree
decomposition}~\cite{Bodlaender93a}; here we recall the equivalent
definition in terms of \emph{elimination orderings}, which, for a graph
$G$, is a permutation $\pi = v_1, \dots, v_n$ of $V(G)$. The
\emph{width} $w_G(\pi)$ of $\pi$ for $G$ is
$\max(\deg_G(v_1), w_{\velim{G}{v_1}}(v_2, \dots, v_n))$, where $\velim{G}{v_1}$ is obtained from
$G$ by removing $v_1$ and completing its neighbors to a clique.
\emph{Treewidth} $\mathrm{tw}(G)$
is the minimum width of an elimination ordering.
Checking if $ tw(G) \leq k $ is NP-complete.

It is well known that if $G$ contains a graph $H$ as minor ($H$ is obtained from $G$ by deleting vertices and edges and by contracting
edges), then $\mathrm{tw}(H) \leq \mathrm{tw}(G)$,
i.e., the class of graphs
of treewidth $\leq k$ is \emph{minor-closed}.
\lv{
A graph without isolated vertices is
\emph{treewidth-critical} if each of its minors has smaller treewidth,
i.e., contracting or deleting any edge decreases the
treewidth. 
}
A famous theorem of \citet{RobertsonS04} states that every
minor-closed family of graphs is definable by a finite set of
forbidden minors, but since the proof is not constructive, finding
these finite \emph{obstruction sets} is an open challenge.
The minimal obstruction set for the class of graphs of treewidth $\leq k$ consists of
\emph{treewidth-critical} graphs; whose any minor has strictly smaller treewidth.
\lv{
	In
particular, all graphs with treewidth $\leq k$ can be characterized by
a finite obstruction set, and it is easy to see that it consists of
exactly all treewidth-critical graphs of width $k + 1$.  
Given an integers $n,k$, the task is to find all
treewidth-critical graphs with $n$ vertices of treewidth $k$.}

\pbDef{arg1}
{Find all treewidth-$k$-critical graphs with $n$ vertices.}
{For showing that a graph's treewidth is at most~$k$, SAT encodings based on elimination orderings are known~\cite{SamerV09}. }
{Negating the encoding by \citeauthor{SamerV09} allows us to express that the treewidth must be at least $k$.}
In principle, one can encode treewidth-criticality by checking that any edge deletion or contraction results in a graph with treewidth $<k$. We opted for a more compact encoding, only requiring that the graph itself has treewidth $\leq k$, and postprocessing to filter out non-critical graphs.

\subsection{Snarks}
\label{sec:snark-def}

A \emph{snark}
is a non-3-edge-colorable cubic graph (\citet{Gardner76} took the name, a portmanteau of `snake' and `shark,' from Lewis Carroll's poem \emph{The Hunting of the Snark}).
Similarly to vertex coloring, edge coloring is NP-hard.
To avoid trivial cases, it is usually required that a snark have girth $\geq 5$ and be \emph{cyclically
$4$-edge-connected} (deleting any $3$ edges does not create two connected components both containing a cycle).

\pbDef{Snarks}
{Enumerate all snarks with $n$ vertices.}
{We use sequential counters to enforce cubicity.
We forbid 3 and 4-cycles by explicitly enumerating them and requiring at least one edge from each to be absent.
Instead of enforcing cyclical $4$-edge-connectedness, we only enforce at least $2$(-vertex)-connectedness.
}
{Universal variables $c_{ij}^l$ for $i<j \in [n]$, $l \in [3]$, and straightforward constraints describe the $3$\hy edge\hy coloring and ensure it is not proper.}
Snarks were introduced in the context of a conjecture now known to be true as the four color theorem: that every planar graph is 4-vertex-colorable.
One equivalent statement of the four color theorem is that planar snarks do not exist.
Snarks continue to be relevant today; for many important conjectures in graph theory~\cite[such as the famous Cycle Double Cover
Conjecture;][]{Szekeres73,Seymour79,Jaeger85} it is known that the smallest
counterexamples, if they exist, must be snarks.  There is already a
wealth of work on enumerating small snarks; for
example, all snarks with up to $36$ vertices are
known~\cite{BrinkmannG17}.

\subsection{Kochen-Specker Graphs}
\label{sec:ks-def}

\emph{Kochen-Specker (KS) vector systems} are special sets of vectors in at least
3-dimensional space that form the basis of the Bell-Kochen-Specker
Theorem, demonstrating quantum mechanics' conflict with classical
models due to
contextuality~\cite{BudroniEtal22}. \citet{KochenSpecker67}  originally
came up with a 3D KS vector system of size~117. The smallest
known system (in 3D) has 31 vectors~\cite{Peres91}, while
the best lower bound is 24 \cite{KirchwegerPeitlSzeider23,LiBrightGanesh23}.  
These lower bounds were obtained with computer search for KS
candidate graphs, which are \emph{non-010-colorable} graphs with additional restrictions.
A graph is 010-colorable if its vertices can be colored red and blue such that no two adjacent vertices are both red and no triangle is all blue.

\pbDef{Kochen-Specker graphs}
{Enumerate all KS candidates with $n$ vertices.}
{For the full list of constraints and the encoding, see~\cite{KirchwegerPeitlSzeider23}.}
{Universal variables $c_v$, $v \in [n]$, and straightforward constraints enforce non-010-colorability.}

\section{Results}
\label{sec:results}

We evaluated all solvers on all benchmark problems, on a cluster of machines with different processors\footnote{
Intel Xeon \{E5540, E5649,  E5-2630 v2,  E5-2640 v4\}@ at most 2.60 GHz, AMD EPYC 7402@2.80GHz}, 
running Ubuntu 18.04 on Linux 4.15.%
\footnote{
	Solvers and benchmark generators are included in supplementary material in the versions used in this paper~\cite{smsqbfaaai25}.
	For detailed instructions on how to create the encodings and use the solvers, as well as for up-to-date version of solvers, visit
	\url{https://sat-modulo-symmetries.readthedocs.io/en/latest/applications#qbf}.
}

\begin{table}
\begin{tabular}{@{}lccc@{}}
\toprule
Task     & CCL & \qbfstatic & \qbfsms   \\
\midrule
$\Delta$-free non-3-colorable & 15 & 13 & 15 \\
Max-$\Delta$-free non-4-colorable & 20 & 12 & 20 \\
Kochen-Specker & 20 & - & 20 \\
Domination conjecture (\emph{i}) & - & 12 & 28 \\
Domination conjecture (\emph{ii}) & - & 12 & 34 \\
Domination conjecture (\emph{iii}) & - & 12 & 30 \\
Folkman $(3,3;4)$ & - & 11 & 15 \\
Folkman $(3,3;5)$ & - & 10 & 12 \\
Treewidth 4 & - & \phantom{0}9 & \phantom{0}9 \\
Treewidth 5 & - & \phantom{0}9 & \phantom{0}9 \\
Snarks girth 5 & - & 14 & 20 \\
\bottomrule
\end{tabular}
\caption{
	The largest number $n$ of vertices solvable with each of the three main approaches, for each benchmark problem, within $4$ hours of CPU time.
	For \qbfsms, the best QBF solver is reported.
	We implemented CCL only for the first three problems.
	Kochen-Specker instances start at $n=15$; \qbfstatic\ did not solve any.
}
\label{table:results}
\end{table}

Table~\ref{table:results} shows the performance of the three main approaches on problem instances from Section~\ref{sec:problems}, with a time limit of 4 hours for each instance. All solvers are run with a single thread.
Table~\ref{table:graphs-solved} shows the number of solved instances by each solver.
`minqbf' is the static encoding of lexicographic minimality, `minqbf-co' is colexicographic minimality using \newSolver{} as the underlying solver.

Note that our experimental setup is intended to compare different approaches, rather to improve on any of the mentioned problems. Making progress would likely require significantly more CPU time and extensive parallelization.

\begin{table}
	\centering
              \setlength\tabcolsep{3mm}
	\begin{tabular}{@{}lr@{\qquad\qquad}lr@{}}
		\toprule
		solver            & \# solved & solver            & \# solved \\
		\midrule
		\newSolver          & 110       & qfun              & 104  \\
		cqesto              &  90       & qute              &  72  \\
		minqbf-co           &  57       & minqbf            &  55  \\
		\bottomrule
	\end{tabular}
	\caption{%
		Number of solved graph instances by each solver.
	}
	\label{table:graphs-solved}
\end{table}

The experiments show that the QBF encoding of minimality performs poorly on almost all instances, independently of the chosen ordering.
Also, Qute does not seem to be well-suited for graph problems. 
Although CCL, CQesto, Qfun, and 2Qiss are conceptually similar, there is variability in performance. 
Among the QBF solvers, \newSolver\ performs best, but CCL, while solving the
same instances of applicable problems, solves them 2--3 times faster.
For the largest solved $n$ from Table~\ref{table:results}, CCL needed 6, 63, and 34 minutes, respectively, while \newSolver\ needed 18, 122, and 58 minutes for the same instances.
For hard combinatorial graph problems requiring multiple CPU years, using CCL might be advisable.

We ran Qute with proof logging with a time limit of 30 minutes per instance (for longer runs, the proofs grow too large) and extracted and validated proofs of solved instances.
Qute solved 70 instances and produced 1874 proofs, which, compressed with \texttt{xz}, total 350MB.
All proofs were verified with \texttt{qrp2rup} and \texttt{gratchk} in under $4$ hours.

\section{Conclusion}
This research explores novel techniques for solving quantified graph search problems with QBF solvers.

Q-SMS, which integrates SMS-style symmetry breaking into QBF solvers, matched CCL performance across multiple prominent graph search problems after a comprehensive evaluation. In contrast, the completely QBF-based method Q-static showed poor scaling beyond small graphs.

Our implementation extends Qute to produce verifiable proofs, a capability missing in CCL approaches. Q-SMS delivers equivalent performance with more straightforward implementation, broader problem applicability, and enables formal proof generation.
Future work will expand \newSolver{} to handle general QBFs and optimize its
core algorithms to improve efficiency on complex graph problems
further.

\section*{Acknowledgements}
This research was funded in  part by the Austrian Science Fund (FWF)
10.55776/COE12 and 10.55776/P36688.
The results were supported by the MEYS within the dedicated program ERC~CZ under the project \emph{POSTMAN} no.~LL1902
and are co-funded by the European Union under the project \emph{ROBOPROX} (reg.~no.~CZ.02.01.01/00/22\_008/0004590).

\bibliography{literature}

\end{document}